# Social Behavior and Mental Health: A Snapshot Survey under COVID-19 Pandemic

Sahraoui Dhelim[1], Liming Luke Chen[2], Huansheng Ning[1], Sajal K Das[3], Chris Nugent[2], Devin Burns[4], Gerard Leavey[5], Dirk Pesch[6] and Eleanor Bantry-White[7].

[1]School of Computing and Communication Engineering, University of Science and Technology Beijing, China.
[2]School of Computing, Ulster University, UK.
[3]Department of Computer Science, Missouri University of Science and Technology, USA.
[4]Department of Psychological Science, Missouri University of Science and Technology, USA.
[5]School of Psychology, Ulster University, UK.
[6]School of Computer Science and IT, University College Cork, Ireland
[7]School of Applied Social Studies, University College Cork, Ireland.

**ABSTRACT**

Online social media provides a channel for monitoring people's social behaviors and their mental distress. Due to the restrictions imposed by COVID-19 people are increasingly using online social networks to express their feelings. Consequently, there is a significant amount of diverse user-generated social media content. However, COVID-19 pandemic has changed the way we live, study, socialize and recreate and this has affected our wellbeing and mental health problems. There are growing researches that leverage online social media analysis to detect and assess user's mental status. In this paper, we survey the literature of social media analysis for mental disorders detection, with a special focus on the studies conducted in the context of COVID-19 during 2020-2021. Firstly, we classify the surveyed studies in terms of feature extraction types, varying from language usage patterns to aesthetic preferences and online behaviors. Secondly, we explore detection methods used for mental disorders detection including machine learning and deep learning detection methods. Finally, we discuss the challenges of mental disorder detection using social media data, including the privacy and ethical concerns, as well as the technical challenges of scaling and deploying such systems at large scales, and discuss the learnt lessons over the last few years.

CCS CONCEPTS • Applied computing →Law, social and behavioral sciences→ Psychology • Computing methodologies → Artificial intelligence → Natural language processing • Computing methodologies → Machine learning → Machine learning algorithms.

**Additional Keywords and Phrases:** Social media analysis, mental disorder detection, COVID-19, mental health.

## 1. INTRODUCTION

Mental illness is the fifth contributor to the global burden of diseases [42]. The economic cost of mental disorders treatment was estimated to be US $2.5 trillion in 2010, and is expected to double by 2030 [16]. One of the main goals of the World Health Organization's (WHO) Comprehensive Mental Health Action Plan 2013–20 was to develop strong information systems for mental wellbeing, including increasing capacity for population health diagnosis [119]. The direct and indirect consequences of COVID-19 on the mental health of the general population were anticipated to be considerable by many commentators. Anxiety, depression and alcohol misuse appeared to have increased in the general population. Grief, loneliness and isolation will play a role in all of these conditions [95]. These impacts may be amplified for already vulnerable groups who rely on other people and organizations for care and support. Moreover, front-line health workers are thought to be vulnerable to burn-out and trauma. There are likely to be longer-term psychological effects of the pandemic created by economic disruption, unemployment and family breakdown [43]. Risk factors such as ill health, bereavement, domestic abuse and violence and maladaptive coping such as increased alcohol consumption, substance misuse and gambling are likely to contribute to increased disorders. The anticipated increase in problems will occur at a time when access to services is severely restricted. The challenges to health and social care service delivery cannot be underestimated and innovative ways are needed to maintain connection to vulnerable people. Online social networks are well established as a data source for public opinion mining [130],

business analytics [38], events detection [32] and population health monitoring [29]; and are increasingly being used for mental health applications, at both population-level and individual-level health. Social media analysis is especially promising for mental healthcare, as Online Social Networks (OSN) such as Twitter and Facebook provide access to naturalistic, first-person accounts of user behavior, emotions, thoughts, and feelings that may be indicative of mental wellbeing. The popularity of social media where people willingly and publicly express their ideas, thoughts, moods, emotions, and feelings, and often share their daily struggles with mental distress offers a rich information source for studying mental illnesses, such as depression and loneliness. In this paper, we survey the existing research literature on social media analysis for mental distress detection, with a special focus on the works in the context of COVID-19 pandemic.

In the recent few years, the number of papers that focused on social media analysis for mental disorders detection has sharply increased, including few survey and review articles. The authors in [126] reviewed the literature on social media analysis for depression and suicide detection, although they limited the review only for text-based social media platforms such as Twitter, Reddit, and Weibo. In [61], 15 papers were summarized to discuss the depressed detection from social media data priory to August 20, 2019. In [72], the authors analyzed the suicidal ideation causes using text-based social media data. Similarly, the authors in [56] reviewed suicidal ideation detection methods using clinical data. However, they covered only a few works of suicidal detection from social media data. With the emerging of COVID-19 pandemic, numerous works have stressed the importance of mental wellbeing during this pandemic, as well as for possible future pandemics. Many new works have used social media analysis as a medium to study the users' mental well-being during COVID-19 pandemic, and most of these works have not been covered in previous reviews and surveys, the main motivation of the current survey is to survey these new works. Moreover, most of the recent surveys and reviews ([126][72][56]) focus only on text-based social media, whereas we cover visual-oriented social media such as Instagram. In Table 1, we list recent surveys and reviews published between 2020 and 2021, and show the difference between these works and the current survey.

Table 1 Recent reviews and surveys on the topic (2020-2021)

| Reference | Scope | Difference with current work |
|---|---|---|
| [126] | Covers only text-based social media. Focuses on depression and suicide ideation detection. | Current survey focuses more on COVID-19 related works |
| [61] | Covers 15 works prior to August 20 2019. Reviews only depression detection. | Current survey covers text-based as well as visual-based social media analysis |
| [72] | Covers suicidal ideation causes analyses from text-based social media data | In addition to depression and suicide ideation, the current survey also covers loneliness, anxiety, stress and post-traumatic stress disorder (PTSD), as well as other mental disorders. |
| [56] | Reviews suicidal ideation detection methods using clinical data, also covers few works of text-based social networks data | |
| [21] | Reviews papers published between 2013 and 2018 | |
| [19] | Reviewed 16 suicide detection techniques using social media data | |

The rest of the paper is organized as follows: In Section 2, we define the scope of the survey and the methodologies considered. Section 3 lists the widely used OSNs and their basic functionalities. In Section 4, we focus on feature extraction from social media content. Section 5 presents mental distress detection techniques. Section 6 reviews the recent mental distress detection literature works and classifies these works according to their feature extraction and detection techniques. In Section 7, we summarize our finding and outline possible future directions. Section 8 presents the open issues and challenges of mental distress detection from social media. Finally, Section 9 concludes the survey.

## 2. SURVEY SCOPE AND METHODOLOGY

The main focus of the current survey is mental disorders, also known as psychological disorders; these two terms are used interchangeably throughout the rest of the paper. Mental disorders incorporate a wide range of mental illnesses, such as depression, anxiety, PTSD and schizophrenia, to name a few. Specifically, we survey the works that use social media data for mental disorders detection. We focus on works published between 2020 and 2021, with a special focus on the works conducted in the context of COVID-19 pandemic. However, we also cover key works from the last few years, such as highly influencer works in the field of mental disorders detection using social media analysis. We adapted PRISMA (Preferred Reporting Items for Systematic Reviews and Meta-Analyses) framework [86] guidelines to select publications related to mental disorders detection using user-generated social media data. As shown in Figure 1, initially, 561 related papers between January 2020 and April 2021 were identified after searching Google Scholar, Elsevier, IEEE Xplore digital library, ACM Digital Library, Springer and PubMed for articles related to the following research queries: "mental disorder", "psychological disorder", "mental distress", "psychological distress", "social media", "social networks", "depression","covid-19", "stress", "anxiety", "loneliness", "PTSD", "suicide", "schizophrenia". The searches were limited to articles written in English. 955 additional articles were identified as related works in the field published between January 2014 and December 2019, these articles were found by following the citations map of the searched articles. After removing duplicated papers, a total of 1516 articles were gathered in the identification phase. In the screening phase, based on the title and abstract screening 980 articles were excluded for not meeting the inclusion criteria. The majority of these articles are either not based on social media data analysis, and based on wearables and smartphones passive sensors, or they focus on non-mental healthcare, such as diabetes, obesity or outbreaks predictions. 416 articles were excluded in the eligibility phase after full-text reading. Finally, 120 articles were qualified for final inclusion.

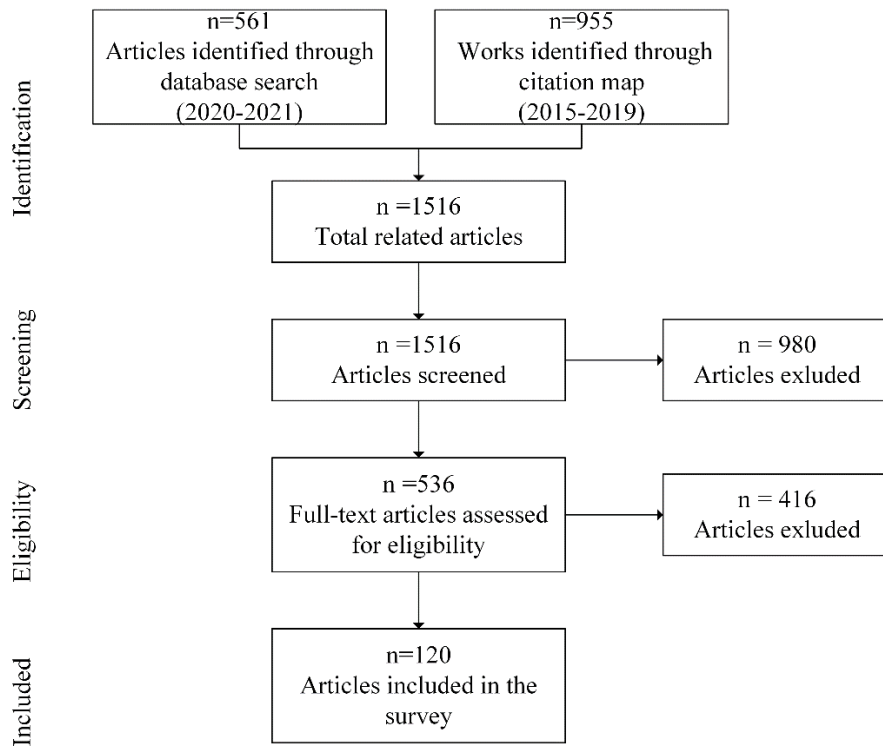

Figure 1 PRISMA flowchart of the review phases

## 3. ONLINE SOCIAL NETWORKS

With the widespread of mobile devices, and the ubiquitous access to the internet, the popularity of OSN had increased sharply during the last few years. Following the travel restrictions and lockdown measures imposed during COVID-19 pandemic, people have turned to socialize using social media, and many OSN have gained more daily active users. These OSN differ in many aspects, some are text-based such as Twitter and Sina Weibo, while other are photo-based such as Instagram and Flicker. The data extracted from these OSN differ in type and features. Table 2 present the most used OSN for mental disorder detection literature. In this section, we introduce the most used OSN in the literature of mental disorder detection, and their data extraction methods.

Table 2 Most used OSN for mental disorder detection

| OSN | Active users in millions (2021) | Main functionality |
|---|---|---|
| Twitter | 353 | Microblogging |
| Reddit | 430 | Forum and blogging |
| Sina Weibo | 511 | Microblogging |
| Facebook | 2740 | Social networking |
| Instagram | 1221 | Photo and video-sharing |

### 3.1. Twitter

Despite its relatively low number of active users (353 million active users by 2021) compared to other social networks. Twitter is by far the most popular OSN in the researcher community. Datasets extracted from Twitter are widely used in the literature of social media analysis. The majority of works surveyed in this work have used Twitter as data source. Every status the user sends to his followers on Twitter is called a tweet. Tweets are publicly accessible and can be extracted and analyzed. In addition to the tweet's text body, tweets also include user metadata, such as user's geographical information and the tweet's date and time, and user networking information. Thus, Twitter data can be used not only for user-level studies, but also for population-level studies. Tweets can be obtained using Twitter API by searching using keywords, hashtags, or other queries, the search can be limited to specific locations or time intervals. In the context of mental distress detection, roughly there are three ways to find and extract the data of mentally distressed users. Some works recruit participants for the experiment and directly extract their Twitter data ([25]), and analyze the extracted data and compare it with their self-assisted scores of various mental healthcare questionnaires. While other works search for keywords or randomly look for mentally distressed users by searching for specific phrases and apply regular expression, for example searching for the phrase "I was diagnosed with depression" or " I feel so down", the downside of this method is that it may require human intervention to confirm the user's mental distress [163][28]. Another way is to search for a specific hashtag to identify tweets related to a particular mental illness, such as #depression or #MyDepressionLooksLike [66].

### 3.2. Reddit

Reddit is a forum-like social network, which consists of thousands of active, user-created communities (also known as subreddits) that focus on particular areas of interest such as art, football, movies, politics, mental disorders and many other topics, it had over 430M active monthly users by 2021. Reddit users can participate in a community by sending text- or media-based posts, and by commenting on other posts or comments. Users can rate posts and comments using up-vote or down-vote functions. The score of a post or comment can be calculated as the difference between up- and down-votes. Many subreddits address mental disorders, including depression, suicide and anxiety, in these subreddits users share their daily life, express their opinions and feelings, and some may offer mental support to others. By March 2021, /r/depression subreddit had 736K members, r/SuicideWatch had 236k members and r/Anxiety had 437k members. Mental health-related subreddits offer a very rich source of data that can be used to train mental distress detection models, since these subreddits contain a high percentage of ground-truth data concentrate in one place, unlike twitter data that require extensive searching to identify such mentally distressed users. Figure 2 shows a recent post of Reddit user asking for help in r/SuicideWatch subreddit.

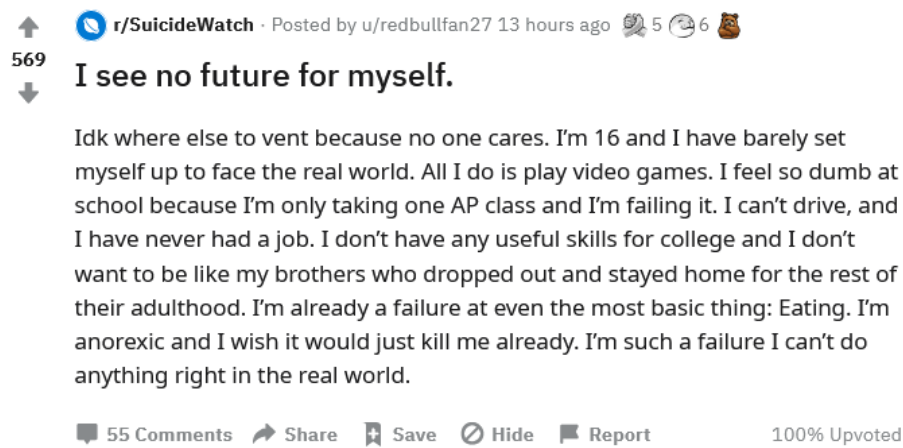

Figure 2 A Reddit user asking for help in r/SuicideWatch subreddit, posted 3 March 2021

### 3.3. Sina Weibo

Sina Weibo is the largest microblogging website in China (simply known as Weibo), its main functionality is very similar to Twitter. Many studies in the literature of mental distress detection were conducted using Weibo datasets. With more than 511 million active users that post about their daily life struggle, Weibo offers a rich source of user mental health information. Various works have collected datasets from Weibo, and these datasets are used for various mental illness detection tasks. Wang et al. [149] extracted the published data of 1M users with more than 390M posts, and employed a keyword-based method to mark users at suicide risk. Following that, three mental health experts manually labeled users at suicide risk. They could identify 114 users with suicide ideation and leveraged linguistic analysis to explore behavioral and demographic characteristics.

### 3.4. Facebook

Facebook by far is the largest social network in terms of the number of active users. Facebook offers the possibility to create and share text posts, as well as photo and videos, users can also join various groups and follow Facebook pages. Unlike microblogging OSN such as Twitter and Weibo, Facebook textual posts are not limited by characters limit (tweets have 280 character limit). In the context of mental distress detection, user-generated data can be used for various purposes, specifically, we can extract textual features, visual features and also behavioral features, see Section 4 for more about the different type of extracted features.

### 3.5. Instagram

Instagram is the largest visual-based social network, where the users share photos and videos. Mental distress detection using media-based OSN may not be as straightforward as text-based OSN, generally, photos and video need deep analysis to associate their features with mental distress markers. Instagram photos incorporate a variety of features that can be analyzed for mental state assessment. The content of photo can be represented by various characteristics: Are there people in the photos? Is the photo setting in outdoor or indoors? The photo was taken night or day? photo statistical properties can also be analyzed at the pixel level, such as average color and brightness. Instagram post metadata contains additional information about the photo: Did the post received any comments? How many 'Likes' did the photo received? Finally, behavioral features, such as usage and posting frequency, may also give some hints about the user's mental state.

## 4. MENTAL FEATURES EXTRACTION

Social network data contains various features associated with mental distress, and could reveal the psychological state of the users. These features are either extracted from a single modalities (e.g., text, images or audio), or fused from a variety of data sources (multimodal) [92]. Multimodal feature processing generally achieves better mental distress detection accuracy compared to single modal schemes, as individuals tend to manifest their inner psychological states using different expression mediums [9]. Hence, multimodal features can alleviate the generalization error caused by individual differences, such as personality type, ethnicity, age and gender. The widespread of mobile computing devices had created a plethora of social networks. Users share their daily life activities; express their opinions and feeling in different social networking platforms. These social networks are used for different purposes; hence, they offer different services centered on different data formats. For instance, text-based microblogging social networks, e.g. Twitter and Weibo, are mainly used for posting short textual messages. While, visual-based social networks, e.g. Facebook and Instagram, are primarily used for sharing photos and videos. Different features are extracted from social network data depending on the type of shared content. We proposed the taxonomy presented in Figure 5 to classify the previous works on mental distress detection using social network data.

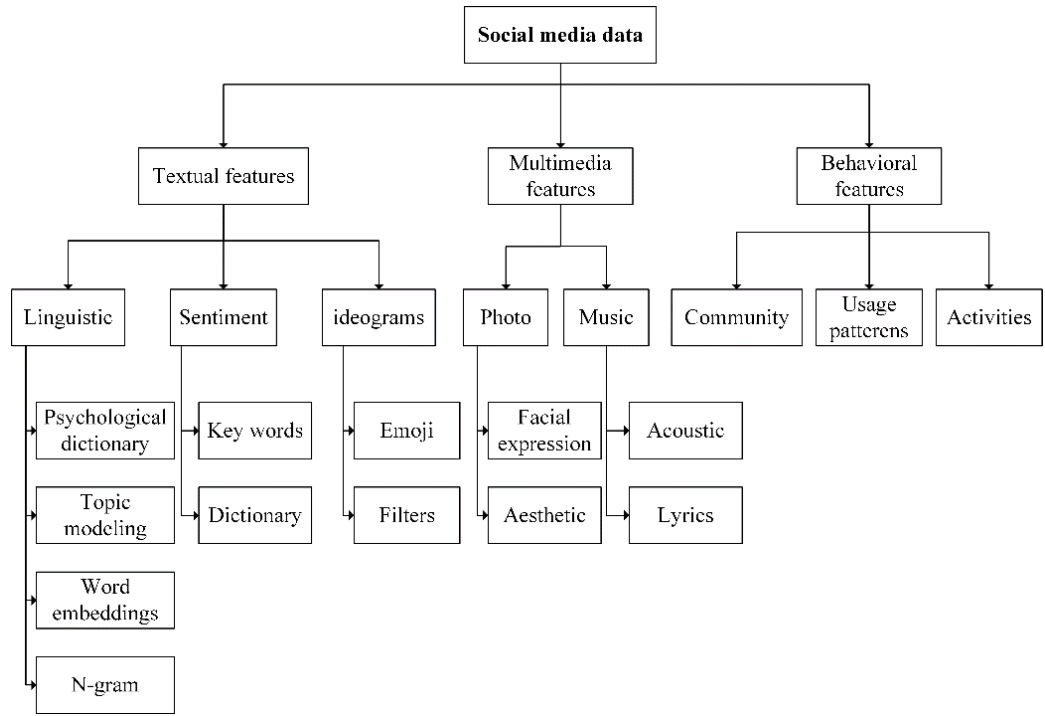

Figure 3 Psychological features classification

## 4.1. Textual features

The user-generated text data incorporate various latent features that can be leveraged to reveal the user's psychological distress. As shown in Figure 3, textual features can be divided into three classes, namely linguistic, sentiment, and ideogram features. Multimodal feature fusing can be applied within the same textual features class by combing various features from the same class (e.g. linguistic-linguistic [134]), or by combining features from various feature classes (e.g. linguistic-sentiment [109]).

### 4.1.1. Linguistic features

The linguistic features are extracted from the language use patterns such as word choice, topics of interest and sentence structure. One of the most used linguistic feature techniques is to map the user-expressed text corpus into a psychological dictionary, and measure the word frequencies in each word class. One of the most widely used psychological dictionaries is Linguistic Inquiry and Word Count (LIWC) [100]. Which is widely used text analysis technique in psychology, and it can be easily adapted for natural language processing (NLP) tasks. LIWC was introduced in the early 1990s to associate linguistic dimensions of written expression with psychological states. LIWC could compute the percentage of words within 80 linguistically or psychologically meaningful categories. These categories cover various important psychological states of an individual, including personal preferences, cognition state and emotions. In the past 30 years, LIWC had been used in different research studying the relationships between the word categories in daily language and psychological states. For instance, the correlation between the first person singular pronoun usage and depression [50], emotions and LIWC emotion-related categories [133], LIWC positive emotions category word usage and anxiety [128]. In the context of mental distress detection using social media analysis, LIWC is by far the most used psychological dictionary-based feature extraction method ([70][51][151][83][167][146]). Other works used other psychological dictionaries to extract linguistic features, such as Affective Norms for English Words (ANEW) [31]. Another effective linguistic feature extraction method is topic modeling, which is the process of extracting the users' topics of interest by analyzing the text corpus generated from their posts. Topic modeling is an effective technique in computational linguistics to decrease the input of textual data feature space to a determined number of topics. Using unsupervised text mining techniques, hidden topics such as topics connected with psychological distress can be extracted from the text corpus.

Unlike psychological dictionaries such as LIWC, it is not created by a fixed set of pre-categorized words. However, topic modeling techniques automatically compute and generate the set of non-labeled words that represent the user's topics of interest. In the context of topic modeling for psychological distress detection, the generated topics are considered the linguistic features of the user. The most used topic modeling method is Latent Dirichlet allocation (LDA) [15]. In LDA, a topic is represented as a multinomial distribution associated to unique words in the text corpus, furthermore, a document is represented as a multinomial distribution over all topics. LDA is used to generate topics and provides linguistic features automatically from the user's text corpus. Most of the work leverage the original unsupervised LDA algorithm for topic extraction ([84][41][112]), some works reported better results using supervised or semi-supervised LDA, that guide the topic modeling process by specifying mental-related lexicon of which are likely to appear in the generated content of distressed individuals ([161][107]).

Another widely used linguistic features for mental distress detection are the word frequency features that represent the original text corpus into representational language model. The most basic models in this category is the Bag of Words (BoW) model [165] and term frequency–inverse document frequency (TF-IDF) [154]. In BoW, the user-generated text is represented as the bag (multiset) of its words, disregarding grammar and even word order but keeping word multiplicity. While TF-IDF is a term-weighting scheme that measure the importance of a word in all text generated by the user, for example TF-IDF might be applied to know the importance of the word "depressed" in all the tweets of a given user. The disadvantage of BoW and TF-IDF is that the order of the words is discarded, and with that we lost important information about the temporal dimension of the user psychological state. The N-gram model [138] solves this problem, N-gram is an adjacent sequence of entities. The entities can be syllables, phonemes, characters, words or base pairs. N-gram models are extensively used in computational linguistics and statistical NLP for various tasks. N-gram model are used as linguistic features that represent the probability of co-occurrence of each input sentence as a unigram and bigram.

Word embedding is a deep learning technique that is used for text representation for the recognition of comparatively important words. This modeling approach is based on the mapping of every word into a corresponding low dimensional vector, where each word is represented as a positive or negative decimal number. One of the widely used word embeddings techniques is Global Vector for word representation (GloVe) [114]. Glove searches for similar words in the whole context of user-generated text. Another widely used word embedding is word2vec [26]. Generally, linguistic features fusing from more than class can increase the mental distress detection accuracy compared to single class linguistic features, e.g. LIWC+N-gram [150], LIWC+LDA+N-gram [134][65].

*4.1.2. Sentiment features*

Sentiment analysis is concerned with extracting emotions, opinions, affects and from user-generated texts on social media, it is crucial that knowledge from this area can also be very useful to find the psychological state of the user. Especially the sentiments that users describe towards their personal situation could be an important marker. Besides the linguistic features described above, the user-generated text on social media could contain semantic features related to mental distress that are difficult to extract using linguistic features extraction, such as emotions and mood. In this context, many works have leveraged sentiment analysis to extract emotional features. The most naïve method of capturing user sentimental features is the key-word search, in which the system looks for specific pre-defined key-words in the users' posts, e.g. 'depressed', 'feeling lonely' or 'happy' [20][122][160]. The drawback of key-word approach is the lack of context of using these words, as user may use these keywords in a sarcastic way for instance. A more selective method is to compute the frequency of used words that belong to LIWC emotion categories ([89][44]), or LDA topics that represent basic emotions ([159]). Another way to measure the importance of emotion-related words is to measure the occurrence frequency of these words using TF-IDF, which is often used in sentiment analysis tasks ([157][3]). In addition to this, the NRC Emotion Lexicon [8] and VADER Sentiment Lexicon [52] are used for user sentiment and emotion recognition.

*4.1.3. Ideogram features*

Most of social networking platforms offer the users the ability to enrich the posted text with additional ideograms such as emojis and stickers, the usage patterns of these ideograms are rich features that can reflect the users' psychological distress. For example, continuous usage of stressed face emoji and stickers might be associated with stress. Emoji usage choices have been associated with depression ([145][80][12]), mental distress [33] and emotions ([139]). One of the main advantages of ideogram feature is the easy and straightforward extraction, unlike linguistic and sentiment features, which require various pre-processing steps.

**4.2. Multimedia features**

With the popularity of visual-oriented social media platforms (e.g. Instagram), and audio streaming services such as (e.g. Spotify), many users tend to use these platforms while being passive on other text-based social networks. Previous researches have proven that visual and audio features are more expressive compared to textual features regarding the psychological states of the users [101].

*4.2.1. Visual features*

Visual features can be extracted from photos and videos posted on social networking websites. Due to the processing difficulties of extracting visual features from videos, most of the previous works focused on feature extraction from photos. There are two types of social media photos, profile photos and posted photos. Generally speaking, profile photos convey less information about the mental state of the user, as most of the user choose socially attractive profile photos, which is known as self-presentation biases, for instance even depressed users express positive emotions in their profile photos [45]. As we have shown in Figure 3, there are three types of features that can be extracted from user-generated photos, facial expression features, object class features and aesthetic features. Facial expression features of the studied users, as well as people present in the photo. Facial expression features not only used to identify the user's sentiments but also information about his exposure to society. For example, the tendency to post group photos is a strong indicator of the social anxiety level of the user. Another indicator is the selfie posting frequency and psychological state [164]. On the other hand, aesthetic features represent the photo characteristics such as the color and filter choices. For example, Photo might be analyzed at the pixel level averages to extract Hue, Saturation, and Value (HSV), three-color properties widely used in image feature extraction. Hue represents the photo's coloring on the light spectrum (varying from red to purple/blue). Lower hue values reflect in more red, and higher hue values result to photo tending more to blue. Saturation represents the vividness of the photo. Low saturation results the photo to be more grey and faded. Value is the image brightness. Lower brightness scores reflect a darker photo. Previous researches associated depression with the posted images being grayscale and low aesthetic cohesion across a variety of image features, and users suffering from social anxiety also tend to post grayscale and low aesthetic cohesion photos but less than depressed users [45], see Figure 4.

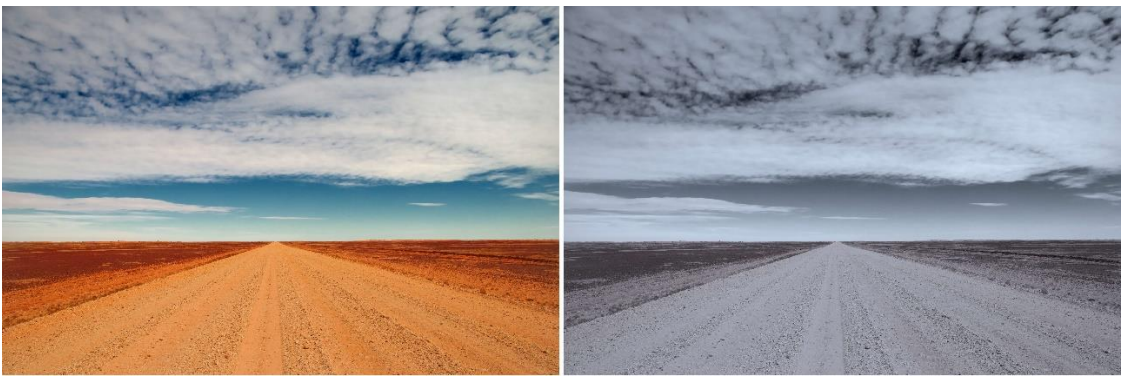

Figure 4 The right photo has higher Hue (bluer), lower Saturation (grayer), and lower Brightness (darker) than left photograph. Instagram photos posted by depressed individuals had HSV values shifted towards those in the right photograph, compared with photos posted by healthy individuals [106].

*4.2.2. Audio features*

Various features can be extracted from music listening activities on online streaming services. Thanks to the available music public dataset and online streaming services API, we can analyze and extract rich features associated with mental distress. By analyzing the user's listening history and playlists, we can extract acoustic features of these songs, such as rhythm structure, timbral texture, pitch, spectral features. Lyrics features are extracted by applying textual feature extraction techniques discussed above on the lyrics of the listened songs. For instance, computing the LIWC categories of the lyrics can be associated with the mood of the users' who listened to this song.

### 4.3. Behavioral features

In addition to the user-generated content feature, the online behaviors convey valuable hints about the users' daily live activities and their interests.

*4.3.1. Communities affiliation*

Social networks offer the user the ability to join social network groups and communities (e.g. Facebook groups, Reddit communities). Analyzing the type of groups and communities that the user have joined might help to detect his topics of interest. Mental distress-related groups and communities are specifically helpful to identify users with mental health issues, many previous studies have shown that topics and psycholinguistic features were found to be accurate predictors of mental disorders like depression [112], bipolar [62] and suicide ideation [121].

*4.3.2. Usage patterns*

The patterns of social media usage and the variation of usage across different periods is a strong indicator of the user's psychological state. Usage features include the distribution of social media usage within the day, the daily usage time and the usage time increase/decrease. For instance, Smith et al. [127] found that the sudden increases of Facebook postings activity is positively correlated with depression symptoms. Lup et al. [75] found that there was a marginal positive correlation between Instagram hourly usage time and depressive symptoms

*4.3.3. Online activities*

Besides the usage time and community affiliations, the interaction activities with posted content, such as likes, comments and follows, might be correlated with certain psychological markers, for example, Negriff [91] found that depressive symptoms is negatively correlated with the number of Facebook friends. Lup et al. [75] investigated the amount of strangers the user follows, and its relation with depression markers, and the number of strangers followed slightly correlated with depression.

### 5. DETECTION TECHNIQUES

Due to the large size and diversity of user-generated data on social networks, many detection models have been used to learn from such rich data. Mental distress detection algorithms are divided into three main classes, machine learning, deep learning and statistical analysis methods. In this section, we survey the various learning models used in the literature of social media mental distress detection. In Figure 5, we present the classification of the detection methods.

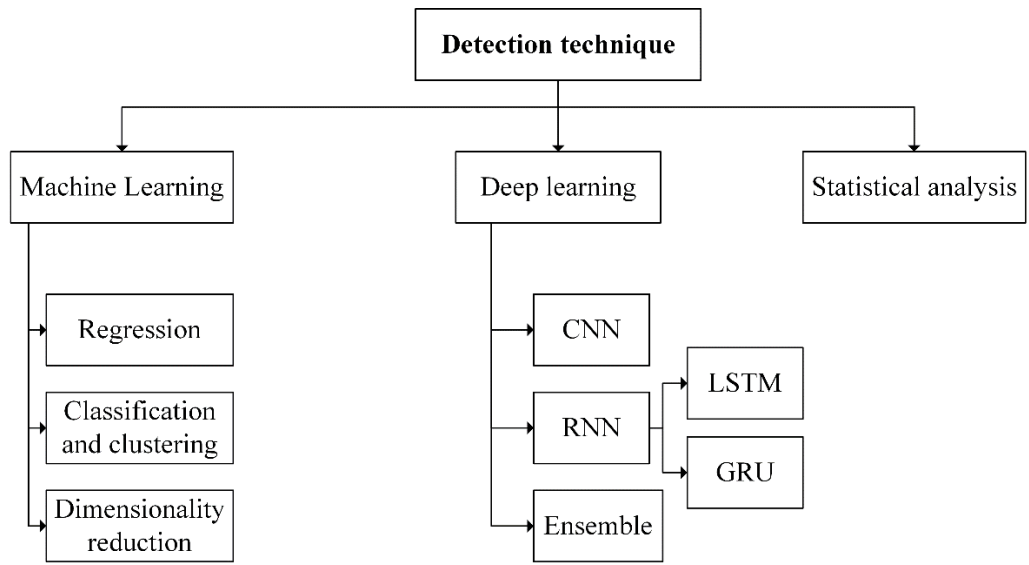

Figure 5 Mental distress detection classification

### 5.1. Machine learning detection

Machine learning detection schemes are by far the most used techniques in the literature of mental distress detection.

*5.1.1. Classification and clustering*

In machine learning, classification is the problem of finding to which of a group/class/category a new observation belongs, based on existing knowledge of a training set of data that includes instances whose class affiliation is already known. For example, given three sets of users grouped based on their depression level, low-group, medium-group and high-group, a classification algorithm is used to analyze the user-generated data on social media and assign the user to one of the three groups (classes). Support-vector machines (SVM) is one of the most effective machine learning classifiers, SVM has been proven to perform well with short informal text, with promising results when applied to mental health classification tasks [25]. SVM is widely used in mental disorders detection using social networks data, including: depression detection ([134][99][25][142]), social anxiety ([22][34][156]), suicide risk assessment ([135][24][76]). Other widely used classifiers for mental health assessments include K-Nearest Neighbor (KNN) [53] and Naïve Bayes (NB) classifier [17].

*5.1.2. Regression methods*

Regression is the process of iteratively refining a measure of the error in the prediction made by the model to find the relationship between variables. Logistic regression is the most used regression model, it is used to predict the risk of mental distress based on extracted social media features of the user, specifically, logistic regression model can calculate the probability of a categorical variable (depressed / non-depressed) from a set of predicted features. Linear regression is also used in various mental disorders detection methods. In linear regression, the relationships between the mental disorder and user data are modeled using linear predictor functions whose unknown model parameters are estimated from the user-generated social media data. For example, in [109] to construct linear regression depression predictive model, linguistic features were obtained from Instagram comments and post captions and, including emoji sentiment analysis results, multiple sentiment scores and meta-features such as likes count and average comment length.

*5.1.3. Dimensionality reduction*

Dealing with raw data in high-dimensional spaces can be undesirable for various reasons; raw data are often sparse as a result of the curse of dimensionality, and analyzing such raw data is usually computationally expensive. Dimensionality reduction is the transformation of raw data from a high-dimensional space into a low-dimensional space in order that the low-dimensional representation preserves some meaningful properties of the raw data, ideally similar to its intrinsic dimension. The objective of dimensionality reduction methods is to learn the inherent latent structure in the data, however, in this case in an unsupervised way to summarize or describe the data using less information. This can be useful to visualize dimensional data or to simplify data, which can then be used later for supervised learning method. Many of dimensionality reduction methods can be leveraged in classification and regression tasks. Principal Component Analysis (PCA) is the most used dimensionality reduction technique in the context of mental disorders assessment using social media data, such as the representation of social data of depressed users [129] and users with schizophrenia [82] and suicidal users [68].

**5.2. Deep learning detection**

In the last few years, Deep-learning methods have become the mainstream detection techniques for mental distress from social networking data. They are focused on building larger and more complex neural networks, deep learning detection schemes are trained with very large datasets of labeled ground truth data, such as linguistic features, social media image, music streaming audio, and video.

*5.2.1. Convolutional Neural Network*

Convolutional neural network (CNN) is a type of deep neural networks, mostly applied to analyzing images. CNNs are special case of multilayer perceptrons (MLP). Multilayer perceptrons typically mean fully connected networks, where every neuron in one layer is connected to all other neurons in the next layer. The "total-connectedness" of MLP makes them prone to overfitting data. This problem is typically solved by regularizing these networks by adding magnitude measurement of weights to the loss function. CNN uses a different method for regularization: they leverage the hierarchical pattern in data and form more complex patterns using smaller and simpler data patterns, making them less connected and complex yet effective. CNNs achieve good results especially when applied in image classification and they generally yield good accuracy for data with a grid-like structure. However, many works studies have proven that CNNs can also be utilized effectively for text classification tasks. Given the user-generated social media data, CNNs can be employed using only a single layer with many filters for sentence classification [141], or more complex multilayer modeled as a sequence of layers that includes an embedding layer, convolutional layer, dense layers, max-pooling layer and the output [62].

*5.2.2. Recurrent Neural Networks*

Recurrent Neural Networks (RNNs) are a class of artificial neural networks in which connections between nodes is in the form of a directed graph along a temporal sequence. Which allows it to model temporal dynamic behavior. Inspired by feedforward neural networks, RNNs utilize their internal state to process variable length sequences of inputs. Which makes them suitable for tasks such as unsegmented, connected series of data. The most famous variants of RNNs is Long Short-Term Memory (LSTM) and Gated Recurrent Unites (GRU). LSTM is augmented by recurrent gates, also known as forget gates, which stop back propagated errors from exploding or vanishing. Instead, errors can flow backward over various virtual layers unfolded in space. This enables LSTM to learn tasks that require memories of events that took place thousands or even millions of discrete time steps earlier. LSTM and its variants such as Bidirectional LSTM (Bi-LSTM) is the dominant deep-learning model in the literature of social media data analysis for mental distress assessment. Some of works opted to use only LSTM or its variants for mental disorders assessment ([55][125][2][49]), while others combine more than one deep learning model such as LSTM+CNN ([155][135][115]) or GRU+CNN [110].

*5.2.3. Statistical analysis*

Besides mental distress detection, some other works have focused on the causal relationship between social media usage activities and mental wellbeing. The correlation analysis allows us to capture the relationships between extracted feature patterns and abnormal habits related to psychological wellbeing. In this class, statistical analysis is widely used to infer any causal correlation or dependence and find the statistical relationship between the user's social media data and his mental wellbeing. For example, correlation analyses were performed in [98] to identify depressive symptom-related features from users of Facebook and Center for Epidemiological Studies-Depression (CES-D) scale scores.

## 6. SOCIAL MEDIA MINING FOR MENTAL HEALTHCARE

In this section, we will review the recent published works on the topic, with a special focus on works that have been conducted in the context of COVID-19 pandemic.

### 6.1. Depression detection

Depression is a common yet serious mental disorder. According to the WHO, more than 264 million people suffer from depression around the world [153]. Symptoms of depression include feeling low self-esteem, hopeless or unhappy, and finding no pleasure in activities you used to enjoy. Depression is different from daily mood swings and short-lasting emotional reactions to everyday life events. Various factors can cause depression such as continuous stressful life events, work pressure, personality disorders, family history and giving birth. Depression could cause a serious health condition, especially with long-lasting moderate or severe intensity. In such case, depression might influence the subject to suffer from acute low mood and being less productive at work or school, or even leads to suicide [36]. Moreover, lockdown measures cause by the spread of COVID-19 has rubbed salt in a wound. Following the economic crises and job lost, add to that the fear of virus that leads the people not to seek mental assessment even when they genuinely needed it, not to mention the feelings of worry, fear, and stress, at individual as well as at community level. The numbers of depressed people had increased sharply since the emerging of COVID-19 [1]. Although there are various effective treatments for depression and other mental disorders, the majority of people in middle-income and low-income countries have no access to mental healthcare [116]. Challenges of effective mental healthcare include a lack of resources, low training quality of healthcare providers and social stigma related to mental disorders. Another challenge to effective mantel healthcare is inaccurate assessment. Even in high-income countries, depressed people usually are not accurately diagnosed, and others with no depression are often misdiagnosed and prescribed antidepressant medication.

#### 6.1.1. Depression assessment

There many depression assessment methods, clinical interviews administrated by practicing clinicians or psychologists are the most reliable method of depression assessment; however, it is extremely difficult to conduct such interviews on a large population [131]. Psychometric self-report questionnaires for depression are generally considered as valid and reliable assessment method. The most prominent depression self-report surveys are the Patient Health Questionnaire (PHQ-9) [64], CES-D [103] and Beck Depression Inventory (BDI) [10]. To the best of our knowledge, only one social media based depression detection study has used clinical interviews to measure the detection accuracy [148]. A few works have asked the users to answer self-report depression questionnaires and used their answers as the ground truth to measure the accuracy of their schemes ([25][105][106]). Alternatively, most of the works used less certain ground truth data, such as the user participating in communities about depression in Reddit ([41][73][54][81]) and LiveJournal [93]. Another method is to mine the shared content on social media to establish ground truth, for example searching the user content to find specific expressions, such as "I was diagnosed with depression" ([28][27][14]), or search for self-report anti-depression medication usage evidence [144].

#### 6.1.2. Depression detection using textual markers

Textual features have been used in many works as depression markers. The textual content of social media posts may also help reveal some signs of depression. For instance, a post that says, "I am feeling down," would be considered as expressing a depressive sign. Monreno et al. [87] analyzed depressive symptoms from Facebook posts of randomly selected university students. They concluded that there was a significant positive correlation between depressive signs expressed in the subject's Facebook post and their score on the PHQ-9 depression scale. Similarly, Ophir et al. [96] study concluded that adolescents who publically expressed daily life stress in their Facebook posts had a higher BDI score than those who did not. In the same vein, Ehrenreich et al. [35] analyzed social media post of adolescent girls, and found that posts contain somatic complaints, negative effect and call for supports are correlated with depression symptoms, while no depressive peers posts were more likely to contain offer of help and support. Shatte et al. [124] have collected 67,796 Reddit posts from 365 fathers, over a 6-month period around the birth of their child. They have used a list of "at-risk" keywords that were suggested by a perinatal mental health expert. postpartum depression was detected by monitoring the change in fathers' use of words indicating depressive symptomatology after their childbirth. In the context of depression related to COVID-19 pandemic, Zhang et al. [166] created an English Twitter depression dataset containing 2,575 distinct identified depressed users with their past tweets, and trained three transformer-based depression classification models based on the collected dataset. Linguistic features have been proved to be good markers of depression. Many works have used psychological dictionaries to analyze language usage and word choice and its association with depression. Shrestha et al. [125] proposed an unsupervised depression detection algorithm based on RNN, they computed a vector of LIWC features for each post text, and feed it to LSTM autoencoder along with network-based features modeling how users connect in the forum. their results on detecting depressed users show that psycho-linguistic features derived from the user's social media posts are good predictors of the user's depression severity. Mustafa et al. [89] used the 14 psychological attributes in LIWC to classify the post into emotions, and assigned weights to each word from happy to unhappy after LIWC classification and trained machine learning classifiers to distinguish the users into three depression levels, High, Medium, and Low. Vanlalawmpuia et al. [143] developed a depression detection method that combines LIWC with another 39 attribute sets and 252 depression-related words with temporal and linguistic style. Tadesse et al. [134] analyzed Reddit users' posts to detect depression from textual features, and found that combining more than textual feature types can yield very high depression detection accuracy. They have combined LIWC, LDA and bigram features, and feed them to MLP classifier resulting in a performance for depression detection of 91% accuracy and 0.93 F1 scores. Ricard et al. [109] have built a depression predictive model that make use of linguistic features, sentiment scores, number of likes, average comment length, and emoji sentiment features.

*6.1.3. Depression detection using multimedia markers*

Depression signs are strongly correlated with the user's multimedia features such as profile photo choose, shared photos and videos, and music listening activities. Compared to other psychological distress and mental disorders, depression has a strong correlation with multimedia features. Reece et al. [106] used CES-D scale as ground truth information, and collected 43,950 Instagram photos of 166 users. They applied color analysis, metadata components, and face detection on the collected photos, and investigated the presence of correlation between the photo properties and depression scores. Specifically, Face detection was used to count the number of human faces in the photo as an indicator for participants' social activity levels. Pixel-level averages were calculated using HSV values, three color properties widely used in photo analysis. They concluded that human ratings of photo attributes (happy, sad, etc.) are weaker predictors of depression, and no correlation was found with the computationally generated features. Similarly, Mann et al. [79] combined various textual features, LIWC, BoW, FastText embedding and ELMo embedding, along with the color of images as visual features. Specifically, they extracted HSV features by taking the average of the pixels in the image, and also counted the number of faces in each image using deep-learning face detection model, resulting in a total of 12 visual features and 64 textual features. Malhotra et al. [78] proposed a multimodal depression prediction model that leverages text, images and videos shared by the user to implement a joint representation. The model uses word2vec to extract textual features, VGG-16 to extract image visual features and Faster R-CNN to extract video visual features. Finally, these features are utilized to obtain weighted average score, which is used for making the final prediction using the Softmax prediction layer.

*6.1.4. Depression detection using behavioral markers*

Online activities can be rich information to study the preferences and behaviors, as well as the differences among users. Smith et al. [127] found that the sudden increases of Facebook postings activity is positively correlated with depression symptoms. Negriff [91] found that depressive symptoms is negatively correlated with the number of Facebook friends. Negriff also concluded that the connectedness of social network (i.e., number of mutual friends) is positively correlated with less depression signs, and depressed people tend to use the location tag function and the 'like' button less often. Reece et al. [105] found that the average word count per tweet was negatively correlated with depression. Nonetheless, unlike [127], they did not find any significant correlation between depression symptoms and the frequency of tweets. Lup et al. [75] studied the relationship between depression signs and Instagram usage time, and amount of strangers the user follows. They found that there was a marginal positive correlation between Instagram hourly usage time and depressive symptoms, and the number of strangers followed slightly moderated this relationship. In other words, depression signs increased with Instagram usage time if the user followed a high number of strangers, but if the user followed fewer strangers, Instagram usage time and depression signs were unrelated.

Table 3 Recent depression detection schemes

| Work | Textual feature | Visual feature | Behavioral feature | Detection technique | Social network |
|---|---|---|---|---|---|
| [134] | LIWC+LDA+bigram | N/A | N/A | MLP | Reddit |
| [125] | LIWC | N/A | Network (follow) | LSTM | ReachOut.com |
| [39] | Word embeddings | N/A | N/A | CNN-BiLSTM | Twitter |
| [89] | LIWC+ TF-IDF | N/A | N/A | Various classifiers | Twitter |
| [124] | Key-words | N/A | Engagement+ community affiliation | SVM | Reddit |
| [81] | TF/IDF | N/A | N/A | Logistic Regression | Reddit |
| [140] | BoW+TF/IDF | N/A | N/A | SVM | Reddit |
| [4] | Word embedding (Glove+ Word2Vect) | N/A | N/A | LSTM | Twitter |
| [2] | Plain text | N/A | N/A | BiLSTM | Twitter |
| [143] | LIWC+keywords | N/A | N/A | Statistical analysis | Facebook |
| [164] | N/A | Aesthetic | N/A | Logistic regression | WeChat |
| [79] | LIWC, BoW, FastText, ELMo | HSV | N/A | LSTM | Instagram |
| [106] | N/A | HSV | N/A | | Instagram |
| [5] | Word embedding (Word2Vec) N-gram (TF/IDF) | N/A | N/A | Various classifiers | Online forums |
| [71] | Raw text | Raw images | N/A | CNN (image) BERT (text) | Twitter |
| [78] | word2vec | VGG-16 Faster-RCNN | N/A | Ensemble (word2vec+ VGG16+ Faster-RCNN) | N/A |
| [137] | Psychological dictionary (Empath+ g Textblo) | N/A | N/A | PCA | Twitter |
| [59] | Sentiment, emotions, personal pronoun, absolutist words, negative words | N/A | N/A | LSTM | Twitter |
| [123] | GloveEmbed, Word2VecEmbed, Fastext and LIWC | N/A | N/A | BiLSTM | Reddit |

### 6.2. Suicide ideation detection

Mental disorders are major risk factor of suicide, according to a WHO report, approximately 800,000 people take their lives every year [152]. The US National Institute of Mental Health classifies suicide into three levels: suicidal ideation, suicide attempts and completed suicide. Suicidal ideation is the desire to commit suicide with no real attempted yet, which is a vital step for suicide risk assessment. Conventional suicidal ideation detection methods mainly depend on clinical assessment or self-reported questionnaires. Many suicidal ideation scales and evaluation tools have been developed, such as Suicide Probability Scale [30], Adult Suicide Ideation Questionnaire [108], Suicidal Affect-Behavior-Cognition Scale [48]. These questionnaires are effective and easy to conduct, but are prone to false negatives due to the participants' deliberate concealment, not to mention, the high cost and difficulties to conduct these questionnaires on a large-scale and over a long-time. The prevalence of social media has presented a unique opportunity a new method for detection suicidal ideation and suicidal cause analysis. People with psychological distress usually do not trust traditional mental health methods and services. Researches showed that people with psychological distress tend to look for help from informal resources, such as social networking platforms instead of seeking psychological expert help [7]. Many researches have confirmed the effectiveness of suicidal ideation detection by analyzing the user's online activities and the generated social data. Sarsam et al. [117] discussed the importance of emotions in Twitter content in detecting suicide risks. They analyzed the features of Twitter users' emotions and behavior responses (sadness, fear, anger, joy, positive, and negative) using SentiStrength and NRC Affect Intensity Lexicon (NRC-AIL) classification. A semi-supervised learning scheme based on Yet Another Two-Stage Idea (YATSI) classifier to identify suicide-related tweets. Schoene et al. [121] studied the possibility of automatic suicide notes identification from social media posts as document-level classification tasks. Specifically, they have extracted suicide related LIWC features extracted from suicide notes and trained a Dilated LSTM suicide risk detection model. Tadesse et al. [135] evaluated the n-gram analysis to prove that the phrases related to suicidal tendencies and low social engagements are often present in suicide-related forums. They studied the transition towards the social ideation related with different psychological states such as heightened self-focused attention, frustration, anxiety, hopelessness, or loneliness. Furthermore, they have extracted various textual features such as TF-IDF and compared the performance of CNN, LSTM and LSTM-CNN combined models, in addition to various machine learning classifiers including SVM, XGBoost, RF and NB. They found that LSTM-CNN combined model significantly outperforms other models achieving 93.8% accuracy and 92.8% F1 score.

Table 4 Recent Suicide ideation detection schemes

| Work | Textual feature | Visual feature | Behavioral feature | Detection technique | Social network |
|---|---|---|---|---|---|
| [121] | LIWC | N/A | N/A | Dilated LSTM | Reddit |
| [117] | SentiStrength, NRC-AIL | N/A | N/A | YATSI | Twitter |
| [135] | N-gram, TF-IDF, BoW | N/A | N/A | LSTM-CNN | Reddit |
| [18] | BERT | Raw images | Posting time | knowledge graph | Weibo Reddit |
| [47] | BERT, Sentence-BERT, GUSE | N/A | N/A | DNN | Reddit |
| [118] | SentenceBERT | N/A | N/A | T-LSTM | Twitter |
| [74] | LDA, NMF | N/A | N/A | CNN | Twitter |
| [97] | Deep Contextualized Word Embeddings (CWE) | N/A | N/A | ANN | Facebook |
| [104] | keywords, VADER sentiment | N/A | N/A | RF | Twitter |
| [111] | keywords | N/A | N/A | neural network | Twitter |
| [102] | TF-IDF, BoW | N/A | N/A | Various classifiers | Twitter |
| [120] | N/A | Image tags | N/A | classifier | Instagram |
| [77] | Word embeddings | visual representation (ResNet) | N/A | GRU | Weibo |
| [67] | SentiWordNet, POS | N/A | Posting time | Statistical analysis | Twitter |
| [90] | TF-IDF, Word2Vec | N/A | N/A | Various classifiers | Vkontakte |

### 6.3. Loneliness detection

The outbreak of COVID-19 has resulted in distressing and unexpected social isolation for many people. Fear from the virus and social distancing rules affected people's mental health, which negatively impacted their feelings, mood, daily habits, and social relationships, which are essential elements of human mental well-being. Specifically, restrictions due to social distancing and quarantines increased feelings of loneliness and social anxiety [113]. Many works have leveraged textual features for loneliness detect from social media content. Xuan and Liew [63] analyzed the patterns of loneliness expression in Twitter during COVID-19 pandemic, and pointed out key areas of loneliness expression across various communities. Specifically, they searched Twitter feeds for tweets that contain 'COVID-19' and 'loneliness' posted between May 1, 2020 and July 1, 2020. Following that they apply topic modeling to extract topics discussed by lonely users, and used Hierarchical Modeling to distinguish overarching topics. Variations in the prevalence of these topics were analyzed over time and across the number of followers of Twitter users. Guntuku et al. [46] selected users whose Twitter posts contained the words 'lonely' or 'alone' and compared these users to a control group selected by gender, age and time of posting. Following that, they filtered the topics and studied patterns of users' posts, and their relation with linguistic features of mental health and studied the effect of language on the prediction of social media manifested loneliness. Similarly, Ameko et al. [6] proposed a loneliness detection system named LonelyText, specifically, they applied SVM classifier on LDA topics that were extracted from Facebook dataset. Multimedia features are also useful in the context of loneliness and social anxiety detection. Kim et al. [60] analyzed the relationships between user loneliness and color features of their Instagram photos. They analyzed 25,394 Instagram photos in terms of color diversity, colorfulness, and color harmony. Their results suggest that color diversity is negatively correlated with user loneliness, in particular romantic loneliness. Behavioral features have proven to carry rich information regarding lonely and social anxious users. Ye et al. [162] studied the relationship between user loneliness and the type of social media they use. 155 Japanese university students involved in the study were divided into four groups based on their response to loneliness questionnaires. Twitter users group, Twitter and Facebook users group, Twitter and Instagram users group and users of all three social media sites. Following that, the effects of social media usage time and usage type on loneliness and well-being for each group were analyzed. They found that, No social media usage has effects on loneliness or mental health were associated for students who used only Twitter or both Twitter/Instagram. Students that use both Twitter/Facebook, loneliness was reduced when they use Twitter and Facebook more frequently but was increased when the students posted more tweets. Students of all three social media were lonelier and had lower levels of mental well-being when they use Facebook via computer longer; however, their access time of Facebook using mobile phone helped them decrease loneliness and improve their levels of mental well-being.

**6.4. Anxiety detection**

A growing number of studies suggesting the number of people with anxiety increased during stay-at-home COVID-19 orders [37]. Chang et al. [22] investigated users suffering from social anxiety disorder using behavioral and social-network topological features. They first collected ground truth data from various social media sources and identify anxious users with the help of mental healthcare professionals. After that, they used multiple features, mainly TF-IDF word scores, negative self-disclosure, sentiment score, parasocial relationship and social event attending. Finally, they applied SVM classifier to identify users with social anxiety disorder. Similarly, Dutta et al. [34] found that when SVM classifier is applied to behavior and interaction features can predict the user's social anxiety disorder status with 79% accuracy and 84% area under the receiver-operating characteristic curve. In the same vein, Fauziah et al. [40] have applied random forest and xgboost classifier on YouTube comments for anxiety detection during COVID -19 pandemics. Xu et al. [158] investigated the relationship between various mental disorders and online activities, they have analyzed Flickr dataset by applying multimodal detection technique that uses multimodal features, namely textual features, visual features such as color distribution and the presence of faces and objects, and metadata features and their relation to mental wellbeing. They found that users suffering from mental distress such as social anxiety have more posting activity during the afternoon and evening as compared to healthy users.

**6.5. Stress detection**

Su et al. [132] investigated the impact of COVID -19 lockdown measures on individuals' psychological states in Italy and China. They firstly extracted data of Wuhan, China Weibo users using the geo-location filter, and the social media data of Lombardy, Italy Twitter users also using geo-location filter. Following that, they extracted all users' posted content with two weeks before and after the lockdown for each region (Wuhan and Lombardy lockdowns). After that, they extracted psycholinguistic features of these posts using the Simplified Chinese version and Italian version of LIWC. Finally, they performed various Wilcoxon tests to study the changes in the psycholinguistic features of the posts before and after the lockdown in Lombardy and Wuhan respectively. Their finding suggests that users focused more on "home", and showed a higher level of cognitive process after a lockdown in both Lombardy and Wuhan. After the lockdown, the stress level decreased, the focus on leisure increased in Lombardy, and the focus on group, religion, and emotions became more common in Wuhan after the lockdown. Wang et al. [147] proposed a three-leveled framework for stress detection. The three-leveled framework learns the personalized stress representations following increasingly detailed processing, i.e., from the generic mass level, group level, to the final individual level. The first mass-level is dedicated on mining the generic stress features extracted from people's linguistic and visual posts using a two-layer attention mechanism, in this layer they used GRU model, which inputs the embedding vector of each word and outputs the hidden representation. The second group-level leverages a Graph Neural Network (GNN) to learn the social media group affiliation features. The third individual-level incorporates the user's personality traits into the proposed stress detection framework. The performance study on 1,324,121 posts collected from the social media accounts of 2,059 Weibo users shows that the proposed framework can achieve over 90% accuracy in detection stressed users. Khan et al. [58] applied LDA on non-stressed related tweets from Twitter to detect stress among the users from their tweets and categorized the tweets into two classes stressed and non-stressed user groups. Their results suggested that applying LDA for stress detection yield better performance than SVM-based detection. Tajuddin et al. [136] proposed a hybrid ontology for stress detection that captures the users' keyword matching search process used in social media to identify the stress-related messages. Meshram et al. [85] applied CNN classifier on linguistic, visual and interaction features extracted from twitter data to identify stressed individuals.

**6.6. Other disorder**

The consequences of COVID -19 pandemic did not only caused mental distress to previously healthy people, but also have aggravated and worsened the situation for people with previous mental issues, such as Schizophrenia and PTSD.

Li et al. [69] analyzed 19,224 schizophrenia-related Weibo posts and extracted psycholinguistic features of the Simplified Chinese version of LIWC from each post. Following this, they applied SVM, NB and logistic model trees (LMT) to identify schizophrenic users. Similarly, Birnbaum et al. [13] collected 3,404,959 Facebook messages and 142,390 images of 223 participants with schizophrenia spectrum disorders (SSD). They analyzed linguistic and visual features uploaded up to 18 months before the first hospitalization using machine learning and built classifiers that identify schizophrenic users from healthy users. Specifically, they used LIWC for linguistic features, and for visual features they have extracted ten features related to hue, nine features related to saturation and value, two measures related to sharpness, one for contrast, in addition to the number of pixels, width, height, and brightness. They applied logistic regression on these features to determine schizophrenic users. Johnson et al. [57] have identified a lexicon of terms that are more common among veterans with PTSD prone to Angry Outburst (AOB) specific pre-crisis data in social media posts. They have collected tweets dataset of general population and veteran with PTSD; they extracted PTSD related topics by searching specific hashtags such as "#PTSD", and analyzed tweets posted by 6,000 veterans. Murarka et al. [88] crawled 17159 Reddit posts during COVID -19 pandemic to identify users with PTSD by analyzing unstructured user data using RoBERTa. Which has a similar architecture to BERT with an improved pre-training procedure.

## 7. FINDINGS AND FUTURE DIRECTIONS

Social media analysis is an effective mental health assessment medium, however, as we have shown in this work, there are various online social networks (OSN) that contain different types of data used to extract various features and processed with different detection techniques. Twitter and Weibo are the most used OSN, due to the short length of microblogging posts that make them suitable for text processing, as they contain the extract of the user's thoughts without any additional wordiness, unlike Facebook posts for instance. The vast majority of the surveyed works used textual features as markers for psychological distress, as textual features are relatively easy to process and transform to machine-readable format. Also because the psychology literature provides strong shreds of evidence about the correlation between language usage and psychological distress. Unlike visual and behavioral features that require huge data input to capture the user's preferences and relate it to psychological patterns, and also the relatively little evidence of the correlation between color preferences for instance, and psychological distress. Machine-learning detection techniques usually yield good performance with a small set of data collected from participants' social media accounts, while deep-learning have the upper hand when dealing with large dataset of the general population generally collected by keywords or hashtags crawling. Specifically, LSTM and its variants are widely used for textual feature based mental distress detection and CNN is mostly used for visual features processing, such as detecting the number of faces in Instagram photos as an indicator of social exposure. Few works have used multimodal data from various OSN, however, they used a relatively small dataset, and therefore the benefits of such an approach are not achieved yet. Developing a multimodal framework that incorporates various user activities is a promising future direction. As different individuals tend to manifest mental status differently in their behaviors, some may manifest their mental status verbally (e.g. linguistic and language usage changes), while other may manifest that aesthetically (e.g. photo colors and filters preferences), and some other may manifest more passive symptoms that can be captured through monitoring their music listening history and online activities.

## 8. OPEN ISSUES AND CHALLENGES

Leveraging people's social media content as a mental healthcare data source for disorders assessments and intervention confers various benefits (i.e. reduced recall bias, cost efficiency, large-scale population-level assessment). However, relying on social media as a data source will pose significant challenges and ethical dilemmas that must be addressed to ensure that such technology is ready for population-level exploration. This section discusses these limitations and challenges.

### 8.1. Privacy concerns

The subjects' privacy is by far the most challenging issue when dealing with social media data, from a research perspective as well as a deployment perspective. From a research point of view, the studied users' privacy might be affected throughout the data processing stages. Although most of the time the studied datasets are publicly available, the problem arises when users' personal attributes can be predicted, and the identity of users can be revealed. Various jurisdictions require certain conditions for researches that can compromise users' privacy. The most common procedure is that the researchers must acquire an ethical approval or exemption from their Institutional Review Board (IRB) prior to the study. In addition to that, they must obtain informed consent from users when possible, protect, and anonymize sensitive data during all research stages. Moreover, they need to be careful when linking data across sites is necessary. Finally, when sharing their data, they need to make sure that other researchers also adhere to the same privacy guidelines [11]. Researchers can rely on public social media datasets for mental healthcare research as long as they ensure the preserving of the confidentiality of users. The privacy concern is more challenging when applying such mental healthcare solutions on the existing social media in large-scale level. As most of the OSN stores the users' data on their servers, and already obtained the users' consents to process their data for service enhancement, however, an additional consent to use their data for mental health purposes is still required.

### 8.2. Detection certainty

Social media based mental healthcare heavily depends on NLP, ML and deep learning. Unfortunately, these computing techniques are not fully mature for deployment in a large scale with sensitive applications like mental assessment. Inaccurate assessment of users with a mental disorder may lead to undesirable consequences, such as false positives where the system falsely detects inexistent mental distress and false-negative where the system misses the detection of mentally distressed users. Although many researches have achieved high accuracy for detecting various psychological conditions (e.g. 91% accuracy detection depression [134] and 93.8% accuracy detecting suicide ideation [135], 90% accuracy detection stress [147]). However, the experiment that yielded these results are more or less in a limited scale, and such high accuracy is not guaranteed when these technologies are applied in a large-scale population. Besides the technological challenges to increase the detection accuracy of these systems, another problem is the credibility of the user-generated content. In many cases, the users portrayed emotions and behaviors on social media are not necessarily a reflection of their actual emotional and psychological status.

### 8.3. Public acceptability

In addition to the privacy concern, the lack of public support and acceptability of using their social media data for monitoring their mental health is yet another challenge. Previous studies show that users have worries that these technologies can be used against them, for example, many users have expressed concerns that mental assessment using their public social media data may negatively affect their credit card or insurance applications or influence their employment career [94]. This fear is backed by the fact that only small portion of the population suffer from mental disorders, however, with the recent lifestyle changes brought along with the spread of COVID-19. People start to release the importance of mental healthcare, now is the best time to convince the public that social media based mental healthcare is needed more than ever before, and to show them the advantages of such approach compared to the conventional mental healthcare, such as non-invasive assessment.

## 9. CONCLUSION

In this work, we have surveyed the field of mental distress detection using social media content, with a special focus on the context of COVID-19 consequences on mental health. We have classified psychological-related features that can be extracted from the social media content and reviewed mental distress detection techniques, including machine learning and deep-learning mental distress prediction models. Finally, we have reviewed recent works on the topic, and classify these works according to their feature extraction and detection techniques. This survey also highlights the challenges of mental disorder detection using social media data, including the privacy and ethical concerns, as well as the technical challenges of scaling and deploying such systems at large scales, and discuss the learnt lessons over the last few years.